# Balance Measures Derived from Insole Sensor Differentiate Prodromal Dementia with Lewy Bodies

Masatomo Kobayashi[1*], Yasunori Yamada[1], Kaoru Shinkawa[1], Miyuki Nemoto[2], Miho Ota[2], Kiyotaka Nemoto[2], Tetsuaki Arai[2]

[1]Digital Health, IBM Research, Tokyo, Japan

[2]Department of Psychiatry, Division of Clinical Medicine, Institute of Medicine, University of Tsukuba, Ibaraki, Japan

*mstm@jp.ibm.com

*Abstract*—Dementia with Lewy bodies is the second most common type of neurodegenerative dementia, and identification at the prodromal stage—i.e., mild cognitive impairment due to Lewy bodies (MCI-LB)—is important for providing appropriate care. However, MCI-LB is often underrecognized because of its diversity in clinical manifestations and similarities with other conditions such as mild cognitive impairment due to Alzheimer's disease (MCI-AD). In this study, we propose a machine learning-based automatic pipeline that helps identify MCI-LB by exploiting balance measures acquired with an insole sensor during a 30-s standing task. An experiment with 98 participants (14 MCI-LB, 38 MCI-AD, 46 cognitively normal) showed that the resultant models could discriminate MCI-LB from the other groups with up to 78.0% accuracy (AUC: 0.681), which was 6.8% better than the accuracy of a reference model based on demographic and clinical neuropsychological measures. Our findings may open up a new approach for timely identification of MCI-LB, enabling better care for patients.

## I. Introduction

Dementia has become an urgent societal issue as aging populations increase, and early detection is necessary to provide appropriate care [1, 2]. Also, different types of dementia, such as Alzheimer's disease and dementia with Lewy bodies, require different types of care, and early differentiation is thus required [3, 4]. Ideally, dementia should be detected and differentiated in its prodromal stage before the onset of disease, which is often called mild cognitive impairment (MCI). Earlier identification enables better preparation for care before clinical symptoms become debilitating [5]. The benefits of earlier identification include not only assisting clinicians in arranging treatment options that are known to be effective for specific types of dementia, but also assisting patients and families in learning how to manage the disease [2]. However, MCI is often overlooked for various reasons including lack of access to clinical resources and misdiagnosis as other diseases [6].

Dementia with Lewy bodies is the second most prevalent form of late-onset neurodegenerative dementia, following Alzheimer's disease, and it is estimated to account for up to 22.8% of all people with dementia [7]. Unfortunately, despite its importance, dementia with Lewy bodies has received less attention than Alzheimer's disease, which has led to lack of recognition and misdiagnosis, and thus to less than ideal care [8]. In addition, this type of dementia is particularly difficult to identify in its MCI stage (i.e., MCI due to Lewy bodies, or MCI-LB) because of the slightness and diversity in its clinical manifestations. For example, it is known that patients with MCI-LB have relatively preserved memory [5]. Also, many patients do not present with parkinsonism, even though it is a core clinical feature of MCI-LB [5]. Therefore, costly and often invasive examinations are required for detection and differentiation in this stage. The clinical diagnostic criteria for MCI-LB require invasive biomarkers and extensive specialists' observation [5], which are difficult to perform during routine check-ups, at primary care, or in daily life. Emerging digital health technologies have the potential to solve this problem by providing noninvasive, easy-to-use tools that assist in MCI-LB identification and can be used in non-specialist settings.

A promising way to assist in MCI-LB identification is the use of sensor-based measurement of postural control capability. Postural control has been well associated with Parkinson's disease [9], another form of Lewy body spectrum disorder [10], along with many other clinical conditions such as arthritis [11] and daily living risks such as falls [12]. The postural control capability can be quantified by balance measures such as pelvis movement [13] and center-of-pressure (CoP) trajectories [14], and these measures can be captured with digital health tools such as wearable accelerometers [13] and insole sensors [15]. Regarding prior work involving dementia with Lewy bodies, Mc Ardle, *et al.* [16] studied accelerometer-based measures captured during a quiet standing task to differentiate dementia subtypes. However, their work did not deal separately with MCI. Leandri, *et al.* [17] reported an increase in CoP sway in patients with Alzheimer's disease and amnestic MCI, but they did not address differentiation between different types of dementia. Overall, knowledge on changes in postural control capability in the MCI stage is limited [18], and there has been little work on differentiation between different dementia types.

The aim of this study was to investigate the usefulness of balance measures captured by an insole sensor to differentiate MCI-LB. To this end, first, we used an insole sensor to collect CoP data during 30 s of quiet standing from 98 older adults comprising patients with MCI-LB or MCI-AD and cognitively normal controls. Then, machine learning and deep learning techniques were applied to the dataset to build and

This work was supported by the Japan Society for the Promotion of Science, KAKENHI (grant 19H01084).

TABLE I. PARTICIPANT DEMOGRAPHICS AND COGNITIVE MEASURES

|  | MCI-LB (n=14) | MCI-AD (n=38) | Cognitively Normal (n=46) | p |
|---|---|---|---|---|
| Sex (% female) | 8 (57.1%) | 17 (44.7%) | 28 (60.9%) | .326 |
| Age, y | 74.6 ± 5.0 | 73.4 ± 4.5 | 70.7 ± 4.9 | **.006** b,c |
| Education, y | 12.3 ± 2.7 | 13.2 ± 2.7 | 13.4 ± 2.2 | .351 |
| Height, cm | 156.6 ± 10.4 | 158.6 ± 7.7 | 157.6 ± 8.4 | .730 |
| Weight, kg | 55.7 ± 9.3 | 57.5 ± 9.0 | 57.7 ± 10.2 | .782 |
| Mini-Mental State Examination | 28.1 ± 1.2 | 26.8 ± 1.7 | 28.1 ± 1.6 | **.001** a,c |
| Clock Drawing Test | 6.9 ± 0.4 | 6.6 ± 0.8 | 6.7 ± 0.9 | .636 |
| Logical Memory IA | 8.1 ± 4.0 | 7.0 ± 3.3 | 11.3 ± 3.4 | **<.001** b,c |
| Logical Memory IIA | 7.1 ± 3.8 | 4.1 ± 2.9 | 9.5 ± 3.2 | **<.001** a,b,c |
| Trail Making Test part A | 51.9 ± 21.9 | 42.1 ± 15.6 | 34.2 ± 10.0 | **<.001** b,c |
| Trail Making Test part B | 167.8 ± 86.6 | 141.2 ± 95.0 | 93.2 ± 50.5 | **.002** b,c |

Values are displayed as mean ± SD and bold values highlight statistically significant differences examined by one-way analysis of variance, except for sex, which is displayed as number (% female) and was examined by a chi-squared test. Significant differences between individual diagnostic groups (chi-squared test, $p < .05$, for sex; Tukey-Kramer test, $p < .05$, for other data) are marked with a, b, or c (a: significant for MCI-LB vs. MCI-AD; b: significant for MCI-LB vs. cognitively normal; c: significant for MCI-AD vs. cognitively normal).

evaluate classification models for discriminating MCI-LB from the other groups.

Our contributions in this study are fourfold. First, we developed a novel, insole-sensor-based pipeline to detect and differentiate MCI-LB from a 30-s quiet standing test. Second, we developed a unique dataset of balance measures involving MCI-LB, MCI-AD, and control participants. Third, we conducted a comparative experiment involving multiple machine and deep learning models. Fourth, we showed that the resultant models could identify MCI-LB with accuracy comparable to that of a reference model based on neuropsychological examinations administered by specialists, and we showed that the accuracy could be further improved with a model combining the insole-based and neuropsychological measures.

## II. METHODS

The study was conducted with the approval of the Ethics Committee, University of Tsukuba Hospital (H29-065), and the written consent of all participants.

### A. Participants

As listed in Table 1, 98 older adults (53 females, 45 males) participated in this study. Of these, 14 and 38 had a diagnosis of MCI-LB and MCI-AD, respectively, while the remaining 46 were cognitively normal controls. Regarding the diagnosis, the MCI-LB group met McKeith, et al.'s clinical diagnosis criteria for probable or possible MCI-LB [5], while MCI-AD was assessed with the National Institute on Aging and Alzheimer's Association (NIA-AA) core clinical criteria for MCI [19] and the AD Neuroimaging Initiative (ADNI) criteria for MCI [20]. We applied the following exclusion criteria: a diagnosis of dementia, a diagnosis of MCI other than MCI-AD or MCI-LB, or any other disease or disability that would interfere with the balance test or neuropsychological examinations described below.

All participants underwent the balance test and the following neuropsychological examinations administered by trained neuropsychologists: the Mini-Mental State Examination [21], the Clock Drawing Test [22], immediate and delayed recall of Logical Memory Story A from the Wechsler Memory Scale-Revised (Logical Memory IA and IIA) [23], and the Trail Making Test parts A and B [24].

### B. Apparatus and Balance Test Procedure

We used an insole-type sensor, OpenGo [15] (Moticon GmbH), for balance measurement. The OpenGo insoles were equipped with 13 capacitive pressure sensors on each side. Each sensor measured the pressure value (7 bits) at 50 Hz along the vertical axis relative to the reference plane. From the sensor outputs, the OpenGo system produced three-dimensional time series data for each side, comprising the mediolateral (ML) and anterior-posterior (AP) coordinates of the CoP and the total force.

Each participant, wearing the same type of indoor shoes with the OpenGo insoles, was required to stand quietly upright on a flat carpeted floor for 30 s with both feet closed and eyes open. We used two pairs of insoles with different sizes (small: 237 × 85 mm; large: 261 × 92 mm), and the participants wore whichever was more comfortable. This procedure was part of a larger study, which will be published elsewhere.

### C. Data Processing and CoP Features

From the 6-dimensional time-series data (3 dimensions for each foot) obtained from the insole sensor, we calculated time series of the CoP coordinates for the entire posture. Specifically, after application of a low-pass filter (fourth-order Butterworth filter with a cutoff of 10 Hz [25]) to each dimension to reduce noise in the original data, the CoP coordinates in each frame were calculated as a weighted average of the left and right CoP coordinates with the total force values as the weights. Fig. 1 shows the CoP trajectories for representative participants. Note that the insole sensor did not provide any information on the relative placement/orientation of the left and right feet. Given that the participants were standing with their feet closed, we assumed that the left and right insoles were placed parallel and 1 cm apart.

Next, the 2-dimensional time series of CoP coordinates was smoothed by a low-pass filter with a cutoff of 5 Hz, and an additional 1-dimensional time series for speed was added. The CoP speed was calculated by using a Savitsky-Golay filter with a polynomial of order 3 and a filter window of length 5, as recommended in [14]. Finally, the resultant time series was divided into segments by a sliding window 150 frames long (3 s) with a 50-frame interval (1 s). Fig. 2 illustrates the overall method for processing the CoP time-series segments.

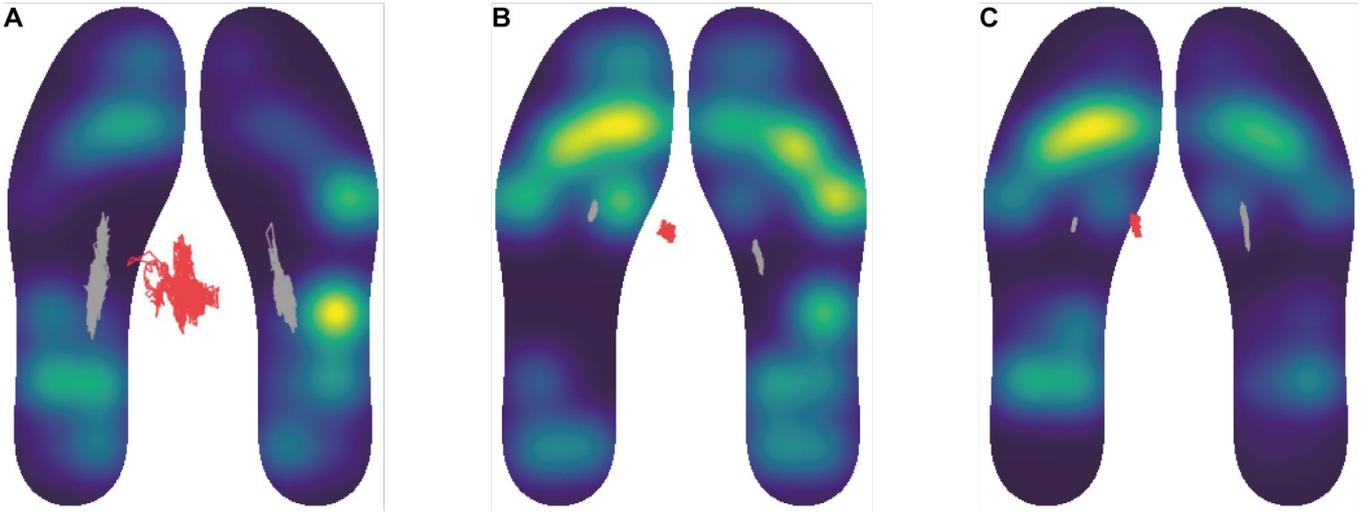

Fig. 1. Representative data for (**A**) MCI-LB, (**B**) MCI-AD, and (**C**) cognitively normal participants. The red and gray trajectries represent center-of-pressure (CoP) movement for the entire posture and for each foot, respectively, while the heatmaps represent pressure mappings calculated from the 13 pressure sensor values for each foot. The MCI-LB patients tended to show larger sway than the MCI-AD or cognitively normal participants in terms of the CoP trajectories.

For each 3-s segment, a total of 18 CoP measures (9 positional, 9 velocity-based) were computed from the 3-dimentional time series of the CoP coordinates (ML and AP) and the CoP speed, in accordance with the literature and a preliminary analysis [14, 26, 27]. The positional measures consisted of the ML and AP ranges, range ratio, root-mean-square ML and AP, ML and AP sample entropies, and ML and AP high-frequency range power ratio. The velocity-based measures consisted of the average CoP speed and its ML and AP components, along with the 10th, 25th, 50th, 75th, and 90th percentiles of the instantaneous CoP speed and its median absolute deviation (MAD). To exclude the effect of outliers, the range was defined between the 2.5th and 97.5th percentiles. Supplementary Table S1 gives the details of each CoP measure.

### D. Machine and Deep Learning Analyses

To assess the capability of machine and deep learning models to discriminate MCI-LB from cognitively normal or MCI-AD cases, we investigated eight supervised classification methods (four machine learning, four deep learning). The machine learning methods we used were LightGBM [28], a support vector machine with a radial basis function kernel, k-nearest neighbors, and logistic regression with L1 regularization. For the deep learning methods, we used Fully Convolutional Network [29], ResNet [29], Time-CNN [30], and Transformer [31], all of which were either developed for or have been applied to time series classification in previous studies. The classification result for an individual participant was obtained by majority voting among the results for segments derived from that participant (Fig. 2).

For the machine learning models, we used the 18 CoP measures for each 3-s segment as input variables. Each variable was standardized by the quantile transformer method and clipped to the range of $[-3, +3]$ to eliminate outliers. We evaluated each model's accuracy through a 10×5 nested cross-validation procedure. Specifically, in the outer loop, the dataset was randomly split into training (9/10) and test (1/10) partitions, and each test partition was untouched while training and only used for performance evaluation. In the inner loop, each training partition was further split into inner training and test folds through another five-fold cross-validation procedure to tune the hyperparameters for the feature selection and binary classification algorithms. We used subject-wise, stratified sampling both in the outer and inner cross-validation procedure, where the participants included in the training and test folds were independent to avoid identity confounding, and each fold contained approximately the same proportion of the two groups. For hyperparameter tuning, we used the tree-structured Parzen estimator method [32] implemented in the Optuna framework [33]. Specifically, the following hyperparameters were tuned in each inner loop: for LightGBM, the strengths of L1 and L2 regularization, maximum number of leaves in one tree, minimal number of data points in one leaf, percentage of features or data points included in each tree, frequency of bagging, and weight of labels with the positive class; for the support vector machine, the strength of regularization, kernel coefficient for the radial basis function, and class weight; for k-nearest neighbors, the number of neighbors; for logistic regression, the strength of regularization and class weight; and for feature selection, the number of features to be selected. The feature selection was performed in accordance with the feature importance in the logistic regression model for the models besides logistic regression, while no feature selection was performed for the logistic regression model. Supplementary Table S2 gives the details of each hyperparameter that we investigated.

For the deep learning models, we used the time series data for each 3-s segment as the input after standardization by the quantile transformer method and clipping to the range of $[-3, +3]$. To enhance the size and quality of the training data, we applied data augmentation techniques such as flipping and jittering [34] to it. Also, prior to the final evaluation, we empirically determined the following hyperparameters on the basis of what values yielded faster and more robust convergence of training losses as well as better performance metrics: for the input data, the data components to be used, direction of flipping, strength and number of jittering applications, and down-sampling of the majority class; for model design, the dropout rate, L1 and L2 regularization, numbers of filters and blocks for Fully Convolutional Network or ResNet, and numbers of heads and blocks for Transformer; for model training, the learning rate and class weight [35]. Using these hyperparameters, we evaluated each model's accuracy through a 10-fold subject-wise stratified cross-validation procedure. In each fold, the model was trained for 250 epochs and the best model was selected in terms of the loss value on the training set. Supplementary Table S3 gives the specific values of each hyperparameter used in the final evaluation.

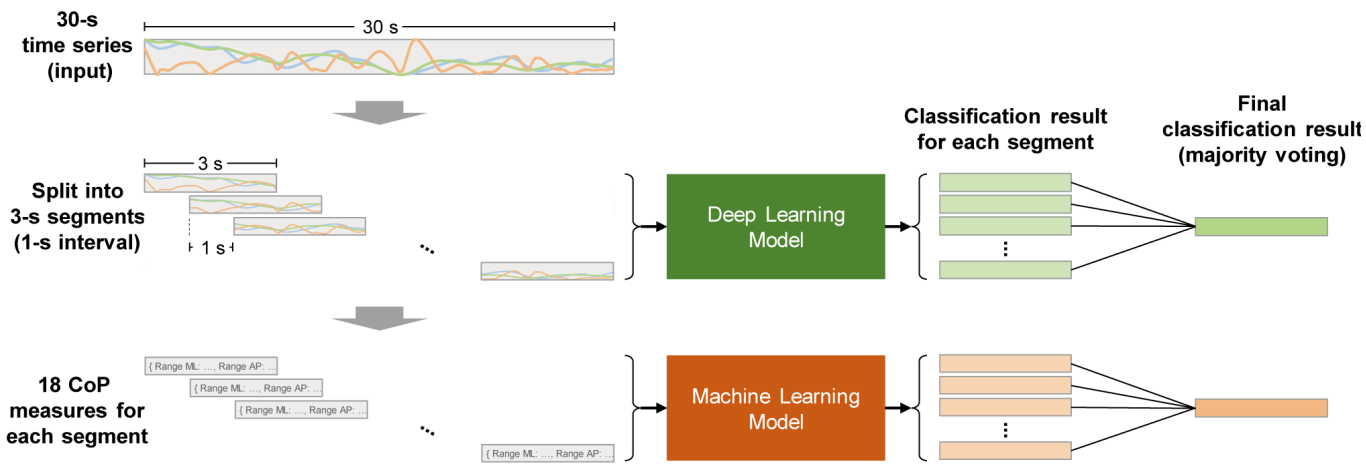

Fig. 2. The experimental workflow. First, the 30-s time-series data for each participant were split into 3-s segments with a 1-s interval. Next, 18 CoP measuers were calculated for each segment. The calculated CoP measues were used by machine learning models for classification, while deep learning models used the raw 3-s time series as input. Finally, the final classification result was determined by aggregating classification results for each segment by majority voting.

For reference, we evaluated logistic regression models that used three demographic variables (sex, age, and years of education) and six neuropsychological measures (Mini-Mental State Examination, Clock Drawing Test, Logical Memory IA and IIA, and Trail Making Test parts A and B) as inputs. We also evaluated machine learning models with hybrid input, which used both the CoP measures and the demographic and neuropsychological measures.

To identify the features (CoP measures) that most contributed to the classification, we further investigated the best machine learning models by using the SHapley Additive exPlanations (SHAP) values [36]. The SHAP values represent the importance of each feature on the basis of their impact on the model output. Specifically, we first excluded non-robust features across different training sets, which were determined as the features with less than 20% occurrences across the cross-validation folds. Then, we evaluated the mean absolute SHAP values of each feature. For the deep learning models, we applied the class activation map [29] method to the models to explore how time series data were used to identify each group, by visualizing which subsequences most contributed to a certain classification.

Finally, to corroborate the analysis results, we conducted statistical analysis on the CoP, demographic, and neuropsychological measures. First, we investigated associations among these measures by using Spearman's rank correlation to examine their dependence and independence. Second, to support the results of the feature importance analysis, we investigated group differences in the CoP measures by one-way analysis of covariance controlling for the age, sex, weight, and height as covariates with the participants as a random factor. The covariates were chosen in accordance with literature reporting the effects of these variables on balance measures [14]. *Post-hoc* pairwise comparisons between the diagnostic groups were performed by using Tukey-Kramer tests.

All analyses were done with the following Python packages: scikit-learn 1.1.3, LightGBM 3.2.1, Optuna 3.0.3, SHAP 0.40.0, and Keras 2.10.0.

### III. RESULTS

Table 2 lists the accuracies for each model. For MCI-LB vs. cognitively normal, the highest accuracy for the models using only insole data was 78.0% (AUC: 0.681), which was 6.8% higher than the reference accuracy based on the demographic and neuropsychological measures. The highest accuracy for the models with hybrid input was 88.1% (AUC: 0.764), which was 16.9% higher than the reference accuracy and 10.2% higher than the accuracy of the best insole-only model. For MCI-LB vs MCI-AD, the highest accuracy for the insole-only models was 74.5% (AUC: 0.545), which was 2.0% lower than the reference accuracy. The highest accuracy for the hybrid-input models was 78.4% (AUC: 0.683), which was 2.0% higher than the reference accuracy and 3.9% higher than the accuracy of the best insole-only model.

To investigate the important CoP measures that most contributed to the machine-learning-based classification, Fig. 3 shows the SHAP values for the best support vector machine models for the features selected in 20% or more of the cross-validation folds. The AP component of the average speed was the most important in both models. For MCI-LB vs. cognitively normal, 10 of the 18 features were included, and four of the top five features in terms of the SHAP values were velocity-based. For MCI-LB vs. MCI-AD, 13 features were included, and all of the top five features were velocity-based. In terms of the direction of measurement, AP measures tended to have larger absolute SHAP values than ML measures.

Next, to investigate how time series data contributed to the deep-learning-based classification, Fig. 4 shows the class activation map for the Fully Convolutional Network model for MCI-LB vs. cognitively normal. It indicates that a subsequence at the beginning with a small variation of the CoP speed and those at the end with highly variable CoP speeds would be the main contributors to the identification of MCI-LB.

TABLE II. MODEL ACCURACIES

| | Input | Model | MCI-LB vs. Cognitively Normal | MCI-LB vs. MCI-AD |
|---|---|---|---|---|
| **Reference** | Demographics; Neuropsychological measures | Logistic Regression | 71.2% | 76.5% |
| **Machine/Deep Learning (insole only)** | Insole data | k-Nearest Neighbors | 74.6% | 72.5% |
| | | LightGBM | 76.3% | **74.5%** |
| | | Logistic Regression | **78.0%** | **74.5%** |
| | | Support Vector Machine | **78.0%** | **74.5%** |
| | | Time-CNN | 68.3% | 61.5% |
| | | Fully Convolutional Network | 70.0% | 67.3% |
| | | ResNet | 75.0% | 71.2% |
| | | Transformer | 76.7% | 71.2% |
| **Machine Learning (hybrid input)** | Insole data; Demographics; Neuropsychological measures | k-Nearest Neighbors | 76.3% | 76.5% |
| | | Logistic Regression | 83.1% | 74.5% |
| | | LightGBM | 84.7% | 76.5% |
| | | Support Vector Machine | **88.1%** | **78.4%** |

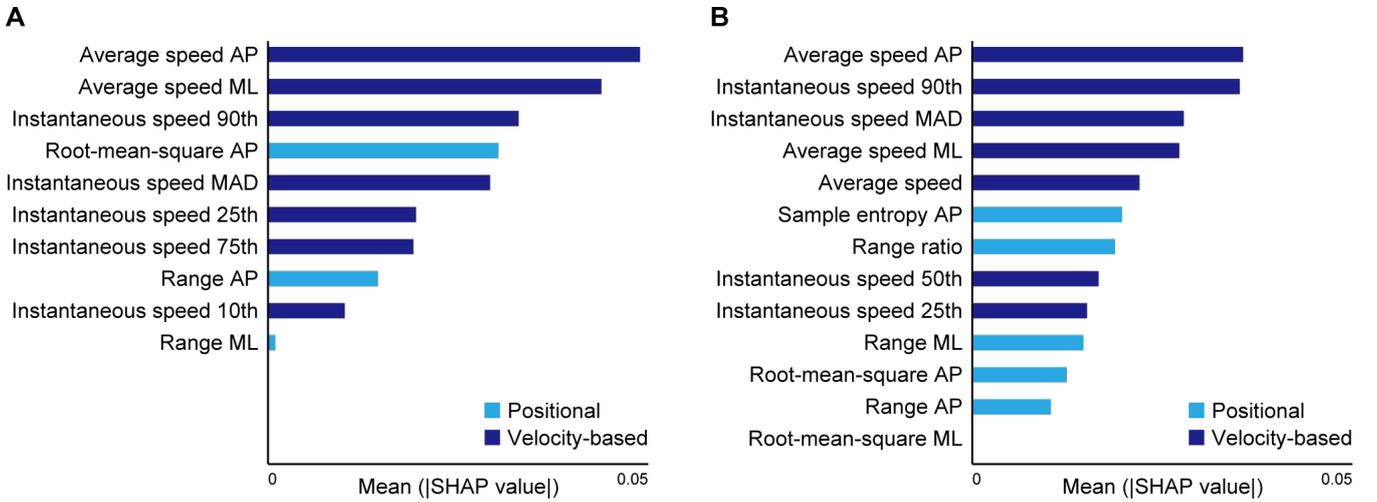

Fig. 3. Comparison of the CoP measures' importance in the classification models for (**A**) MCI-LB vs. cognitively normal and (**B**) MCI-LB vs. MCI-AD. Each plot presents the mean absolute SHAP value for the features selected in 20% or more of the cross-validation folds.

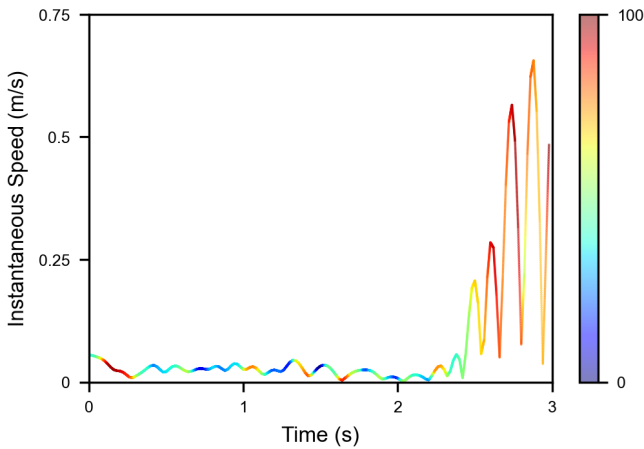

Fig. 4. Class activation map for the Fully Convolutional Network model for MCI-LB vs. cognitively normal. The waveform represents one participant's speed profile. A value of "100" (red) indicates subsequences that contributed to the identification of MCI-LB, whereas "0" (blue) indicates subsequences that did not contribute to the classification.

As mentioned above, the insole-based models performed better for MCI-LB vs. cognitively normal, whereas the reference model performed better for MCI-LB vs. MCI-AD. To corroborate these contradictory results, we examined the correlations within and between the demographic, neuropsychological, and CoP measures. Any correlations between the neuropsychological and CoP measures were very weak (absolute Spearman's rank correlation coefficient: $|\rho| < 0.186$; Fig. 5), whereas there were very strong correlations within the neuropsychological measures ($|\rho| < 0.852$) and within the CoP measures ($|\rho| < 0.987$). The correlations between the demographic and neuropsychological/CoP measures were moderate ($|\rho| < 0.484$ for neuropsychological, $|\rho| < 0.556$ for CoP).

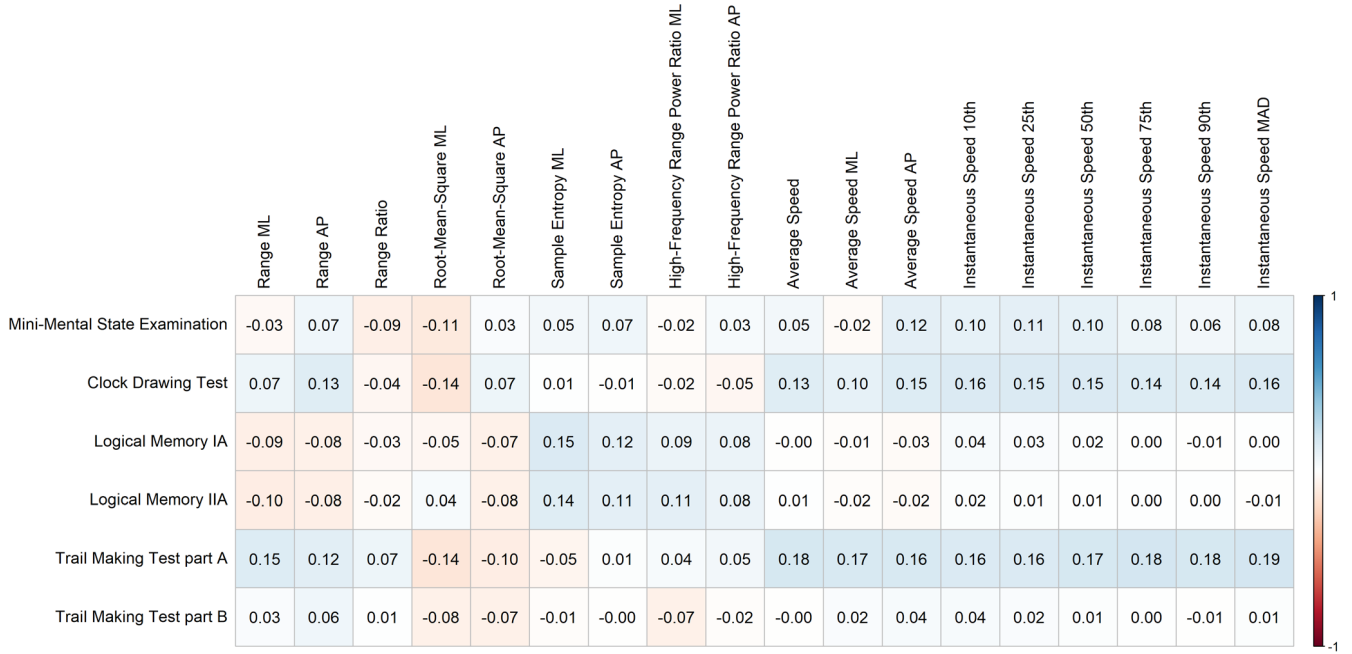

Fig. 5. Correlations between the neuropsychological and CoP measures in terms of the Spearman's rank correlation coefficient, $\rho$.

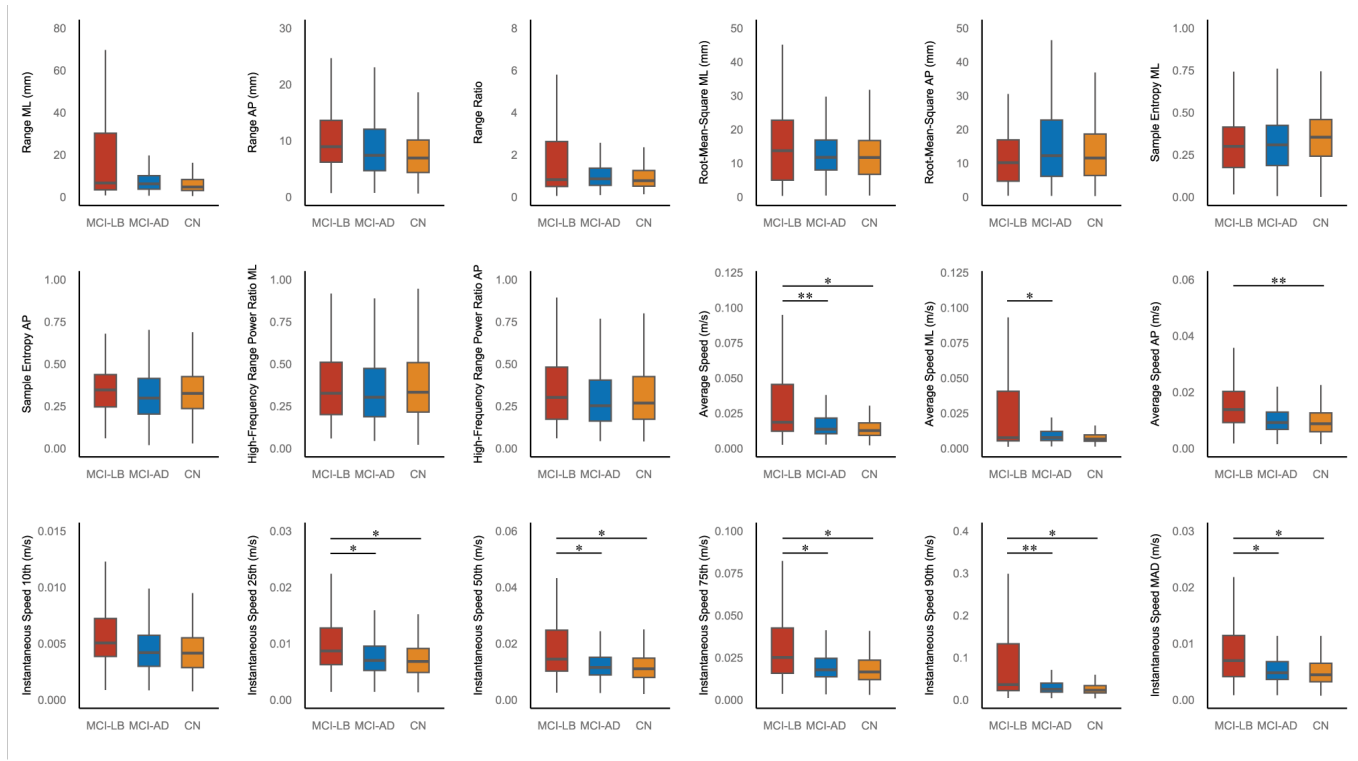

Fig. 6. Group comparisons of the 18 CoP measures. Boxes indicate the 25th (Q1) and 75th (Q3) percentiles; whiskers indicate the upper and lower adjacent values that are most extreme within Q3+1.5(Q3−Q1) and Q1−1.5(Q3−Q1), respectively. Statistically significant differences between individual diagnostic groups are marked with ** ($p < .01$) or * ($p < .05$). MCI-LB: mild cognitive impairment due to Lewy bodies, MCI-AD: mild cognitive impairment due to Alzheimer's disease, CN: cognitively normal.

Finally, we examined the statistical group differences among the 18 CoP features (Fig. 6). Eight of the nine velocity-based features showed statistically significant differences for either or both MCI-LB vs. cognitively normal and MCI-LB vs. MCI-AD, whereas no positional features showed statistical differences. Also, none of the 18 features was significant for MCI-AD vs. cognitively normal.

## IV. DISCUSSION AND CONCLUSION

In the experiment, the proposed machine and deep learning models based on CoP measures derived from an insole sensor could discriminate between MCI-LB and the other groups with accuracies that were comparable to the reference accuracy based on neuropsychological examinations administered by professional neuropsychiatrists. The accuracies could be further improved when the models used both insole and neuropsychological data in combination. Among the investigated CoP measures, the velocity-based ones were more discriminative than the positional ones, and the AP measures tended to be more discriminative than the ML measures. These results were corroborated by additional statistical analyses.

Notably, the machine and deep learning models with only insole data as input achieved higher accuracies than the reference model in discriminating MCI-LB from cognitively normal. In contrast, the benefits from the insole data were relatively small for discrimination of MCI-LB from MCI-AD. These contrasting results were consistent with the statistical analysis, which revealed significant differences in the CoP measures between MCI-LB and the cognitively normal controls, whereas the differences in the neuropsychological measures were relatively small for the two groups. This could be explained by previous findings in the literature that patients with dementia with Lewy bodies tend to have more severe deficits in visuospatial/constructional and motor abilities than patients with Alzheimer's disease [37], while static balance is controlled in a complex way involving different sensory (e.g., vision and tactile) and neuromotor systems [14].

Our experimental results suggest that the CoP and neuropsychological measures capture different but complementary aspects of functional deficits related to MCI-LB. In fact, the correlation between the CoP and neuropsychological measures was weak in our dataset, indicating that the two types of data represent different factors; yet their hybrid use improved the discriminative performance, indicating that they are complementary. In practice, the proposed insole-based system could be used in combination with other digital health tools such as tablet-based drawing tests [38] or linguistic/paralinguistic analysis of spontaneous speech [39]. These behavioral analysis tools have been well associated with neuropsychological measures, and behavioral characteristics derived from different modalities [40] or from different tasks [41] have been associated with different but complementary aspects of cognitive impairments. Therefore, a hybrid use of two different tools could facilitate accurate identification of MCI-LB by complementarily capturing different aspects of the target disease.

CoP measures can be captured with various sensors. In clinical research, expensive professional equipment with a large footprint, such as force plates and stabilometers, has been used; however, that approach is limited to laboratory settings. On the other hand, in the context of digital health, small wearable sensors such as insoles are available. In addition, estimation of CoP trajectories through human pose tracking with video camera images [42, 43] or wearable inertial sensors [42, 44] is becoming feasible. These technologies provide means for continuously monitoring postural control capability outside laboratory settings, i.e., in everyday life situations. This advantage is especially important given that dementia has been associated not only with postural control deficits in quiet standing but also with deficits in the gait and in other activities of daily living [45].

We must acknowledge several limitations of this study. First, our dataset was small and imbalanced, and our findings need to be validated with a larger, more balanced dataset. Second, we evaluated machine and deep learning models only for discriminating MCI-LB from MCI-AD or cognitively normal controls; in clinical practice, differentiation between MCI-AD and controls as well as between MCI-LB from other conditions such as Parkinson's disease is also needed. Third, we determined the hyperparameters of the deep learning models outside the cross-validation loop, which might have led to unreliable results.

In conclusion, we proposed a machine learning-based automatic pipeline to detect and differentiate MCI-LB by exploiting balance measures captured with an insole sensor. The experimental results showed that the resultant models could identify MCI-LB with performance comparable to that of a reference model based on clinical neuropsychological measures. The insole-based models were particularly beneficial for discriminating MCI-LB and cognitively normal, for which the neuropsychological measures did not work well. The results are promising because the proposed procedure involves very simple instructions (e.g., "stand quietly for 30 seconds with your legs closed") and automated sensor measurements, whereas neuropsychological examinations require specialists' administration and more than 30 minutes of time. Our findings may open up a new approach for timely identification of MCI-LB, enabling better care for patients.

Supplementary Table S1.  Center of Pressure (CoP) Measures

| CoP Measure, unit | Description |
|---|---|
| **Positional** | |
| Range ML, mm | Range between 2.5th and 97.5th percentiles of CoP coordinates along the ML or AP axis |
| Range AP, mm | |
| Range ratio | Range ML / Range AP |
| Root-mean-square ML, mm | Root-mean-square of CoP coordinates along the ML or AP axis |
| Root-mean-square AP, mm | |
| Sample entropy ML | Regularity of the CoP trajectory |
| Sample entropy AP | |
| High-frequency range power ratio ML | Power distribution in a high-frequency band (1-5Hz) |
| High-frequency range power ratio AP | |
| **Velocity-based** | |
| Average speed, m/s | Total length of the CoP trajectory / duration |
| Average speed ML, m/s | Total length of the CoP trajectory along the ML or AP axis / duration |
| Average speed AP, m/s | |
| Instantaneous speed 10th, m/s | |
| Instantaneous speed 25th, m/s | |
| Instantaneous speed 50th, m/s | 10th, 25th, 50th, 75th, and 90th percentiles of the instantaneous speed |
| Instantaneous speed 75th, m/s | |
| Instantaneous speed 90th, m/s | |
| Instantaneous speed MAD, m/s | Median absolute deviation (MAD) of the instantaneous speed |

Supplementary Table S2.  Hyperparameters for Machine Learning Models

| Hyperparameter | Name in Python Package | Range |
|---|---|---|
| **LightGBM** | | |
| Strength of L1 regularization | lambda_l1 | $1\times10^{-1}$ to $1\times10^{2}$ |
| Strength of L2 regularization | lambda_l2 | $1\times10^{-1}$ to $1\times10^{2}$ |
| Maximum number of leaves in one tree | num_leaves | 5 to 25 |
| Minimal number of data points in one leaf | min_data_in_leaf | 4 to 16 |
| Percentage of features included in each tree | feature_fraction | 0.2 to 0.6 |
| Percentage of data points included in each tree | bagging_fraction | 0.2 to 1.0 |
| Frequency of bagging | bagging_freq | 1 to 3 |
| Weight of labels with the positive class | scale_pos_weight | 1 to 4 |
| **Support Vector Machine** | | |
| Strength of regularization | C | $1\times10^{-10}$ to $1\times10^{0}$ |
| Kernel coefficient for the radial basis function | gamma | $1\times10^{-6}$ to $1\times10^{0}$ |
| Class weight | class_weight | uniform or "balanced" |
| **k-Nearest Neighbors** | | |
| Number of neighbors | n_neighbors | 1 to 20 |
| **Logistic Regression** | | |
| Strength of regularization | C | $1\times10^{-2}$ to $1\times10^{4}$ |
| Class weight | class_weight | uniform or "balanced" |
| **(Feature Selection)** | | |
| Number of features to be selected | | 2 to 8 |

Supplementary Table S3.  Hyperparameters for Deep Learning Models

| Hyperparameter | Value |
|---|---|
| **(Input Data)** | |
| Data components to be used | Speed component |
| Direction of flipping | Horizontal |
| Strength jittering | $0.03\sigma$ |
| Number of jittering | 3 times |
| Down-sampling of the majority class | Random down-sampling to a ratio of 1:1 |
| **(Model Design)** | |
| Dropout rate | 80% |
| L1 and L2 regularization | None |
| Numbers of blocks (Fully Convolutional Network) | 3 |
| Numbers of filters (Fully Convolutional Network) | 128, 256, 128 |
| Numbers of blocks (ResNet) | 3 |
| Numbers of filters (ResNet) | 64, 128, 128 |
| Numbers of heads (Transformer) | 4 |
| Numbers of blocks (Transformer) | 4 |
| **(Model Training)** | |
| Learning rate | 0.001 (halved each time the loss value showed no improvement for 50 epochs) |
| Class weight | "balanced" (adjusted inversely proportional to the class frequencies in the training data) |